\documentclass[twocolumn,showpacs,preprintnumbers,amsmath,amssymb]{revtex4}
\usepackage{epsfig, graphicx, dcolumn, bm, times, verbatim}

\begin{document}

\title{Effects of degree distribution in mutual synchronization of neural networks}

\author{Sheng-Jun Wang,$^1$ Xin-Jian Xu,$^2$ Zhi-Xi Wu,$^1$ and Ying-Hai Wang$^{1,}$\footnote{Electronic address: yhwang@lzu.edu.cn}}

\affiliation{$^1$Institute of Theoretical Physics, Lanzhou
University, Lanzhou Gansu 730000, China\\
$^{2}$Departamento de F\'{i}sica da Universidade de Aveiro,
3810-193 Aveiro, Portugal}

\date{\today}

\begin{abstract}
We study the effects of the degree distribution in mutual
synchronization of two-layer neural networks. We carry out three
coupling strategies: large-large coupling, random coupling, and
small-small coupling. By computer simulations and analytical
methods, we find that couplings between nodes with large degree
play an important role in the synchronization. For large-large
coupling, less couplings are needed for inducing synchronization
for both random and scale-free networks. For random coupling,
cutting couplings between nodes with large degree is very
efficient for preventing neural systems from synchronization,
especially when subnetworks are scale-free.
\end{abstract}


\maketitle

Synchronization phenomena in neural systems have attracted much
attention. These phenomena are thought to be important for
functioning neural system, such as neural coding, visual
information processing, sleeping, and memory in brain \cite
{Larson, Steriade, Gray, Aihara}. Besides the complete
synchronization, which results in identical states of all neurons
in a uniform population, more subtle forms of synchronization
should be present for examining the brain functions. The brain is
essentially a system of interacting neural networks and the
activity pattern of different networks may become synchronized
while retaining their complex spatiotemporal dynamics. Therefore,
it is interesting to investigate mutual synchronization in
ensembles of coupled neural networks \cite {Morelli}. This problem
has been studied on fully connected networks \cite {Zanette} and
on random networks \cite {Li}.

Recently, it was suggested that connectivity in neural systems is
more complex \cite {Eguiluz}. The effects of complicated
topologies on network dynamics have been confirmed in some
theoretical studies. For instance, small-world neural networks
give rise to fast system response with coherent oscillations \cite
{Lago}. Scale-free Hopfield networks can recognize blurred pattern
efficiently \cite {Stauffer}. Complex networks have both
sensitivity and robustness as responding to different stimuli
\cite {Bar-Yam}. On the other hand, synchronization is not always
desired in neural systems. For instance, several neurological
diseases such as Parkinson's disease and epilepsy are caused by
synchronized firing of neural oscillators \cite {Sakaguchi}. So it
is also important to study desynchronization and instability of
synchronized motion of neural systems.

In this paper, we investigate the efficiency of scale-free
topology in inducing synchronization and preventing the system
from synchronization in two-layer neural networks. We carry out
three coupling strategies between subnetworks: large-large
coupling (couplings built between nodes with large degree); random
coupling (couplings built between randomly selected nodes);
small-small coupling (couplings built between nodes with small
degree). Computer simulations reveal that couplings between nodes
with large degree play an important role either in inducing
synchronization or preventing the system from synchronization. An
analytical treatment confirms the numerical simulation result.

We consider a neural network model that consists of $N$ neurons
$x_i(t) \in (-1,1)$, $i=1,\cdots,N$. The topology of networks was
represented by symmetric adjacency matrix $A$ whose entry $a_{ij}$
$(i,j=1,\cdots,N)$ is equal to $1$ when neuron $i$ connects to
neuron $j$, and zero otherwise. Each link has a weight $J_{ij}$
which is a random number distributed uniformly in the interval
between $-1$ and $1$. The system considered is composed of two
identical neural networks, and a part of corresponding nodes in
different subsystems is coupled together. The dynamics of the
system is described by the following equations \cite {Zanette,
Li},
\begin{eqnarray}\label{eq:model}
&x_i^1(t+1)=(1-{\varepsilon}_i)\Theta(h_i^1(t))+{\varepsilon}_i\Theta(h_i^1(t)+h_i^2(t)),
\nonumber \\
&x_i^2(t+1)=(1-{\varepsilon}_i)\Theta(h_i^2(t))+{\varepsilon}_i\Theta(h_i^1(t)+h_i^2(t)).
\end{eqnarray}
In the equation, $\varepsilon_i$ represents the coupling strength
between the nodes $i$ in different networks. When a pair of nodes
is coupled, the coupling strength between them is equal to a
constant $\varepsilon_i=\varepsilon$, otherwise $\varepsilon_i=0$.
For large-large (small-small) coupling we choose a group of nodes
which consists of the nodes of greatest (smallest) degree in a
subnetwork, and take $\varepsilon_i=\varepsilon$ if node $i$ is in
the group.
Here $h_i^l(t)$ is the local field of the $i$th neuron and is
expressed by
\begin{equation}\label{eq:localfield}
h_i^l=\sum_{j=1}^N a_{ij}J_{ij}x_j^l(t), \ l=1,2
\end{equation}
denoting the signal arriving at neuron $i$ at time $t$ from
neurons of the same network. The local field is determined by the
network topology. The non-coupled nodes in each subnetwork
interact indirectly with another subnetworks through the local
field. The activity function $\Theta(r)$ is defined by
\begin{equation}\label{eq:activity}
\Theta(r)=[1+\tanh({\beta}r)]/2,
\end{equation}
where $\beta=1/T$ characterizes a measure of the inverse magnitude
of the amount of noise affecting this neuron, performing the role
of reciprocal temperature in analogy to thermodynamic systems. For
convenience, we set $\beta=10$ through all simulations. The
initial states of all neurons in two subsystems are randomly
chosen. To measure coherence in this coupled system, the
dispersion of activity patterns is introduced
\begin{equation}\label{eq:dispersion}
D(t)=\frac{1}{2}\sum_{l=1}^{2}\sum_{i=1}^{N} [x_i^l(t)-\overline
x_i(t)]^2,
\end{equation}
where $\overline x_i(t)=\sum_{l=1}^2 x_i^l(t)/2$ is the average
activity of neurons occupying the position $i$ in both subnetworks
at time $t$. The dispersion vanishes when the system reaches the
completely synchronous state.

\begin{figure}
\centerline{\epsfxsize=9cm \epsffile{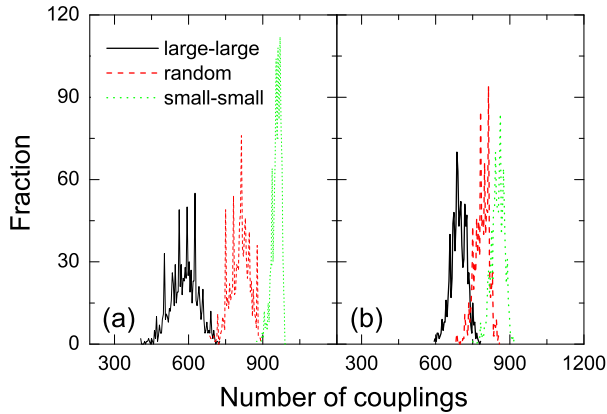}}

\caption{\label{fig:histogram}(color online) Histograms of the
number of couplings needed by the system to reach synchronization
for three coupling strategies: large-large coupling (solid),
random coupling (dash), and small-small coupling (dot),
respectively. The system is composed of BA networks (a) or ER
networks (b) with $N=1000$, $\langle k \rangle=20$, and
$\varepsilon=0.6$. $1000$ simulations were performed for each
histogram.}
\end{figure}

Firstly, we investigate the number of couplings needed between two
subnetworks for system synchronization to study the efficiency of
the network topology. In Fig. \ref{fig:histogram}, we plot the
histograms of the number of couplings built to guarantee
synchronization for scale-free and random networks. We use
Barab\'asi-Albert (BA) arithmetic \cite {BA} to generate the
scale-free network and use Erd\"{o}s-R\'{e}nyi (ER) arithmetic
\cite {ER} to construct the random network. According to the
parameter setting in \cite{Bar-Yam}, we also choose the size
($N=1000$) and average degree ($\langle k \rangle=20$) for both
kinds of networks. When subnetworks are scale-free (see Fig.
\ref{fig:histogram}(a)), the mean fraction of couplings are
$58.2\%$, $80.7\%$, and $95.6\%$, corresponding to the large-large
coupling, random coupling, and small-small coupling, respectively.
This implies that the large-large coupling strategy is more
efficient than the random coupling and the small-small coupling is
the most inefficient method for inducing synchronization. In
addition, the small-small coupling can be regarded as removing
couplings which link nodes with larger degree from the globally
coupled system. Therefore, removing couplings among nodes with
larger degree at the initial state can efficiently prevent the
system from synchronizing. In contrast to the case of scale-free
topology, the peaks corresponding to three coupling methods for
the system consisted of random networks are more closer (see Fig.
\ref{fig:histogram}(b)), which is caused by the homogeneous
distribution of network connectivity. This implies that the
topology of subnetworks and the degree of coupled nodes influence
the efficiency of the system, and the scale-free topology is more
efficient than random network for inducing synchronization or
preventing the system from synchronized states.

\begin{figure}
\centerline{\epsfxsize=9cm \epsffile{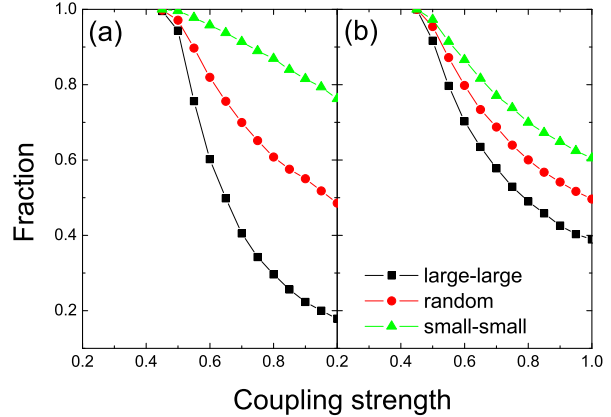}}

\caption{\label{fig:frac-epsilon}(color online) The number of
couplings (fractions of total nodes) needed for reaching
synchronization as a function of the coupling strength
$\varepsilon$ for three coupling methods. The system is made up of
BA networks (a) or ER networks (b). Parameters: $N=1000, \langle k
\rangle=20$.}
\end{figure}

Figure \ref{fig:frac-epsilon} shows the fraction of couplings
needed to induce synchronization versus the coupling strength
$\varepsilon$. Whether subnetworks are scale-free or random, there
is a critical point $\varepsilon_{c}=0.44$ below which partial
coupled networks can not synchronize. For
$\varepsilon>\varepsilon_c$, degrees of nodes taking part in
interactions between two subnetworks will efficiently affect
synchronization or the prevention of networks from
synchronization. The larger the parameter $\varepsilon$ is, the
less couplings are needed for system synchronization. Furthermore,
given the coupling strength, the scale-free topology is more
efficient than the random graph to synchronize in the case of
large-large coupling.

\begin{figure}
\centerline{\epsfxsize=9cm \epsffile{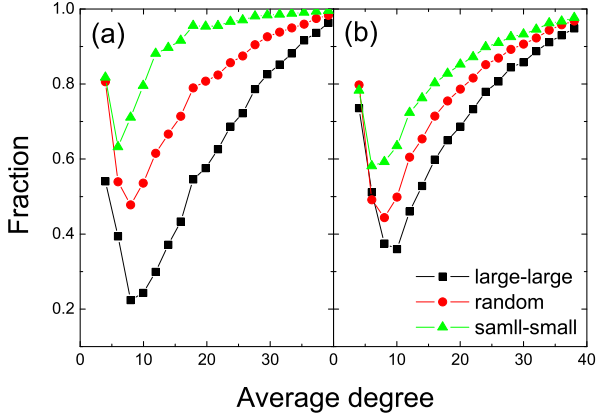}}

\caption{\label{fig:frac-degree}(color online) The dependence of
the fraction of couplings needed for reaching synchronization on
the average degree of subnetworks, BA networks (a) or ER networks
(b). Parameters: $N=1000, \varepsilon=0.6$.}
\end{figure}

We now examine the dependence of the fraction of couplings needed
for synchronization on the average links per nodes $\langle
k\rangle$ in a subnetwork. For small $\langle k\rangle$, due to
few neighbors per node and weak indirect interaction between
non-coupled nodes, more couplings are needed for networks to
synchronize. For large $\langle k\rangle$, different from
classical results of network synchronization \cite {kuramoto}, the
coupled system still needs many nodes to be coupled to ensure
synchronization. Figure \ref{fig:frac-degree} displays a
surprisingly nonmonotonic dependence of the fraction on $\langle
k\rangle$ for given network size and the coupling strength, i.e.,
there exists an optimal value of $\langle k\rangle$ for which the
fraction of couplings reaches minimum. With the increase of
$\langle k \rangle$, the interaction of nodes inside a subnetwork
is enhanced through the local field. The enhanced inside
interaction brings the big indirect interaction between
non-coupled nodes, which helps to the synchronization of
non-coupled nodes. So less couplings are needed for system
synchronization when $\langle k \rangle$ increases. On the other
hand, the evolution of subsystems with stronger inside interaction
is more stable and it needs more couplings to drive their
evolutions into synchronization \cite{Li}. The indirect
interaction between non-coupled nodes and the stability of
subnetworks get a proper match, when $\langle k \rangle$ takes the
optimal value.

Next, we will demonstrate the difference between the efficiency of
scale-free and random networks for inducing or preventing
synchronization. To determine the fraction of couplings needed for
synchronization, we consider a node which does not take part in
direct interaction between two subnetworks. The state of this node
is determined by its local field, $h_i(t)$, defined by Eq.
(\ref{eq:localfield}). The local field includes two types of
signals, one comes from neighbors which interact with nodes of
another subnetwork and the other comes from the rest neighbors.
Thus the total degree of the coupled nodes determines whether the
system can reach the synchronized state through the intensity of
signals in the mean local field.
For the the six schemes investigated in Fig. \ref{fig:histogram},
we calculated the critical values $k_c$ of the total degree of the
coupled nodes when synchronization occurs. These $k_c$ are
normalized by the sum of the degree of the subnetwork $N*\langle k
\rangle$. The standard deviation of the critical values is 0.019,
while the mean of critical total degree of coupled nodes is 0.785.
So
we argue that synchronization will occur if the sum of the degrees
of coupled nodes in one subnetwork exceeds a threshold.

For convenience of notation, the number of coupled nodes is
labelled to be $F_r$ for random coupling and $F_l$ for large-large
coupling. When the system is randomly coupled by either scale-free
or random subnetworks, the average degree of coupled nodes is
nearly the same as the average degree of the subnetworks $\langle
k \rangle$. So the total degree of coupled nodes is
\begin{equation}\label{eq:deg-ran}
\sum_{i=1}^{F_r}k_{i}=\langle k \rangle F_r,
\end{equation}
where $k_i$ is the degree of neuron $i$.
For any network topology with the same average degree and size,
the fraction of couplings is the same in the case of random
coupling. Following ideas developed by Bar-Yam and Epstein
\cite{Bar-Yam}, we get the relation between the fraction of
couplings for random coupling (denoted by $f_r$, $f_{r}=F_{r}/N$)
and the fraction of couplings for large-large coupling (denoted by
$f_l$, $f_{l}=F_{l}/N$).

When subnetworks are random, the total degree of coupled nodes is
\begin{align}\label{eq:deg-lar-ER}
\sum_{i=1}^{F_l}k_i & = N\sum_{k_l}^{\infty}\frac{k \langle k
\rangle ^k e^{-\langle k \rangle }}{k!}
\end{align}
for large-large coupling, where $k_l$ denotes the minimum degree
of the coupled nodes. Thus the fraction of couplings is
\begin{equation}\label{eq:fl-ER}
f_l=\sum_{k_l}^{\infty}\frac{\langle k \rangle ^k e^{-\langle k
\rangle}}{k!}
\end{equation}
So the difference of the fraction of couplings for two coupled
methods is
\begin{equation}\label{eq:relation-ER}
f_r-f_l=\frac{\langle k \rangle ^{k_l-1}e^{-\langle k
\rangle}}{(k_l-1)!},
\end{equation}
which maximum over $k_l$ is obtained approximately by setting
$k_l=\langle k \rangle +1/2$ for a given value $\langle k \rangle$
\cite{Bar-Yam}.

When subnetworks are scale-free, the degree distribution has a
power law shape $P(k)=Ak^{-\gamma}$. The total degree of coupled
nodes is
\newcommand{\ud}{\mathrm{d}}
\begin{align}\label{eq:deg-lar-BA}
\sum_{i=1}^{F_l}k_i
& = N\int_{k_l}^{\infty} kP(k) \ud k \notag \\
& = \frac{1}{\gamma-2}NAk_{l}^{2-\gamma}
\end{align}
with the ancillary condition
\begin{equation}
F_l=N \int_{k_l}^{\infty} P(k) \ud k = \frac{1}{\gamma -
1}NAk_l^{1-\gamma}
\end{equation}
for large-large coupling. Normalizing the probability distribution
and assuming a sharp cutoff of the distribution at low $k$, we
yield
\begin{equation}\label{eq:parameter_A}
A=\frac{(\gamma-2)^{(\gamma-1)}}{(\gamma-1)^{(\gamma-2)}} \langle
k \rangle ^{(\gamma-1)}.
\end{equation}
Combining Eqs. (\ref{eq:deg-lar-ER}), (\ref{eq:deg-lar-BA}) and
(\ref{eq:parameter_A}), the relation between $f_l$ and $f_r$ is
obtained
\begin{equation}\label{eq:relation-BA}
f_l=f_r^{(\gamma-1)/(\gamma-2)}.
\end{equation}
For BA model, the degree exponent $\gamma$ is equal to $3$
\cite{BA}. Thus we have $f_l=f_r^2$.

There also exists a threshold of degree for preventing the system
from synchronization when we cut couplings from the globally
coupled system. This threshold is equal to the difference between
$N \langle k \rangle$ and the threshold of degree for ensuring
synchronization. Similarly, the relationship between the fraction
of removed couplings for large-large cutting and random cutting
also follow Eqs. (\ref{eq:relation-ER}) and
(\ref{eq:relation-BA}).

\begin{figure}
\centerline{\epsfxsize=9cm \epsffile{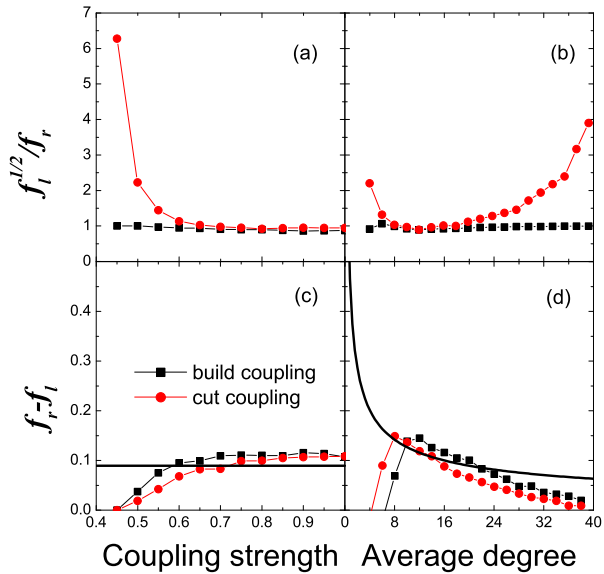}}

\caption{\label{fig:analytic} (color online) Comparison numerical
results with analytical predictions for building couplings to
induce synchronization (square) or cutting couplings to prevent
the system from synchronization (circle). Upper panel: the ratio
$f_l^{1/2}/f_r$ for scale-free subnetworks as of function of the
coupling strength (a) and of the average degree of subsystems (b).
Lower panel: the difference $f_r-f_l$ for random subnetworks as a
function of the coupling strength (c) and of the average degree of
subsystems (d). The thick lines are the maximum of $f_r-f_l$
obtained by analytical calculations.}
\end{figure}

Figure \ref{fig:analytic}(a) shows  $f_l^{1/2}/f_r$ as a function
of coupling strength when subnetworks are scale-free. In the case
of building couplings, numerical simulations give that
$f_l^{1/2}/f_r=1$, which is consistent with Eq.
(\ref{eq:relation-BA}). In the case of cutting couplings, the
ratio is larger than the analytical prediction when the coupling
strength $\varepsilon$ is small, which results from the fat tail
\cite {Dorogovtsev} of BA networks. In other words, the number of
nodes with large degree is more than that described by the
power-law distribution. As a result, some extra couplings between
large degree nodes are removed in simulations and therefore the
number of cut couplings predicted by analysis is less than that of
simulations. When coupling strength is strong, the number of
couplings cut from the coupled system is large, and the simulation
results close to the analytical prediction. %
Figure \ref{fig:analytic}(b) shows the ratio $f_l^{1/2}/f_r$ as a
function of the average degree $\langle k \rangle$ of scale-free
subsystems. For building couplings, simulation results agree well
with the analytical prediction. For cutting couplings, the
simulation results are greater than the analytical calculation in
the case of either $\langle k \rangle$ is low or large. The
deviation results from the small fraction of couplings in these
regions as shown in Fig. \ref{fig:frac-degree}(a). When
subnetworks are random, the difference $f_r-f_l$ are shown in
Figs. \ref{fig:analytic}(c) and \ref{fig:analytic}(d) as a
function of the coupling strength $\varepsilon$ and the average
degree of subnetworks $\langle k \rangle$. The analytical result
of the upper boundary of $f_r-f_l$ gives a good limitation to
numerical results. Although building large-large couplings and
removing large couplings improve the efficiency in inducing and
preventing synchronization, the analytical result of random
subnetworks restricts the enhancement of efficiency to a small
range which is less than that of scale-free subnetworks.

In summary, we have studied the influences of the degree
distribution of networks on mutual synchronization in a two-layer
neural networks. We investigated three coupling methods between
two subsystems: large-large coupling, random coupling, and
small-small coupling. We found that couplings between nodes with
large degree nodes play an important role in the synchronization.
For large-large coupling, less couplings are needed for inducing
synchronization for both random and scale-free networks. For
random coupling, cutting couplings between nodes with large degree
is very efficient for preventing neural systems from
synchronization, especially when subnetworks are scale-free. By
assuming that the total degree of coupled nodes in subnetworks
determines the system synchronization, the numerical simulation
results are interpreted analytically. The analysis reveals that
the degree distribution of subnetworks rather than other
topological quantities affects the efficiency of systems in
synchronization. Although our work is based on a simple model of
neural systems, we think that the results found out in this work
is proper in more wide and realistic situations in which the
dynamics of neurons depend on the mean local field. It would be
interesting if Nature takes advantage of the efficiency of the
scale-free topology in controlling mutual synchronization of
interacted systems.

This work was supported by the Fundamental Research Fund for
Physics and Mathematics of Lanzhou University under Grant No.
Lzu05008.

\end{document}